
\magnification=\magstep1
\def\Buildrel#1\over#2{\mathrel{\mathop{\kern0pt #1}\limits_{#2}}}

\line{ }
\vskip1cm
\centerline{\bf TIME (A-)SYMMETRY IN A RECOLLAPSING QUANTUM UNIVERSE}
\medskip
\centerline{H. D. Zeh}
\centerline{Institut f\"ur Theoretische Physik, Universit\"at Heidelberg}
\centerline{Philosophenweg 19, D69120 Heidelberg, Germany}
\bigskip
\bigskip
\centerline{Contribution presented at Mazagon, Spain, in October 1991}
\bigskip
\centerline{to be published in ``Physical Origins of Time Asymmetry''}
\centerline{ ed. by J.J. Halliwell, J.P.Perez-Mercador and W.H. Zurek}
\centerline { Cambridge University Press 1994}
\bigskip
\centerline{Figures on request from AS3 at vm.urz.uni-heidelberg.de}
\bigskip
{\narrower\smallskip\noindent
{\bf Abstract:} It is argued that Hawking's `greatest mistake' may not have
been a mistake at all.  According to the canonical quantum theory of
gravity for Friedmann type universes, any time arrows of general nature
can only be correlated with that of the expansion. For recollapsing universes
this seems to be facilitated in part by quantum effects close to their maximum
size. Because of the resulting thermodynamical symmetry between expansion
and (formal) collapse, black holes must formally become `white' during the
collapse phase (while physically only expansion of the universe and black holes
can be observed). It is conjectured that the quantum universe remains
completely
singularity-free in this way (except for the homogeneous singularity) if an
appropriate boundary condition for the wave function is able to exclude {\it
past} singularities (as is often assumed).\smallskip}
\bigskip\noindent
{\bf 1 Conditioned entropy in quantum cosmology}
\medskip
Invariance under reparametrizations of time may be considered as a
specific consequence of Mach's principle (which requires the absence of any
preferred or `absolute' time parameter). In quantum theory this leads to a
time-independent Schr\"odinger equation (Hamiltonian constraint), since any
parametrization of physical time (or `clocks') would require the concept of
a trajectory in configuration space. For example, in canonical quantum gravity
the wave function of the universe is dynamically described by the `stationary'
Wheeler-DeWitt equation $H \Psi_{universe} = 0$ in superspace (the
configuration
space of geometry and matter). The conventional time dependence has
then to be replaced by the resulting quantum correlations between all
dynamical variables of the universe including those describing physical clocks,
in particular the spatial metric (see Page and Wootters, 1983). However, this
procedure leaves open the problem of how to formulate the asymmetry in time
which
is manifest in most observed phenomena.
 \par
For example, entropy as the thermodynamical measure of time asymmetry is
defined
in quantum theory as a functional of the density matrix $\rho$,
 $$
S = -k Trace \{ \hat P\rho \ln (\hat P\rho) \} \quad . \eqno (1)
$$
This definition requires an appropriate `relevance concept' or `generalized
coarse graining' which is represented by a `Zwanzig projection' $\hat P$ (an
idempotent operator on the space of density matrices -- cf. Zeh, 1992). Well
known examples of relevance concepts are Boltzmann's neglect of particle
correlations, or the neglect of all long-range correlations (quantum and
classical) in  the form of replacing the density matrix by a direct product
$\hat
P_{local}\rho :=  \Pi_i\rho_{\Delta V_i}$ of density matrices
$\rho_{\Delta V_i}$ for
separate volume elements $\Delta V_i$, each of them obtained from $\rho$ by
tracing out the rest of the world (the `environment'). The latter procedure
gives
rise to the usually presumed local concept of an entropy {\it density}. Under
an
appropriate Zwanzig projection, the density matrix in (1) may even represent a
pure (`real') state, $\rho = |\psi><\psi|$, which should, however, depend on
some
time variable in order to allow the entropy to grow.

Physical entropy, in contrast to the entropy of
information, is objectively defined as a function of macroscopic variables
(such as those characterizing density, volume, shape,
position or temperature), regardless of whether they are known. Therefore,
the wave function $\psi$ to be used in (1) cannot be identified with
$\Psi_{universe}$ (which is a superposition of macroscopically different
states), but must instead represent some `relative state' (conditioned wave
function) for the microscopic degrees of freedom with respect to `given'
macroscopic variables of the universe (including clocks). In the framework of
a global quantum description, this state is understood as the `present
collapse component' (or as `our Everett branch') that has resulted
indeterministically from all measurements or measurement-like processes of the
{\it past}. While the unitary part of von Neumann's dynamical description of
measurements leads to a superposition of macroscopic `pointer
positions', its components can be considered as dynamically decoupled from one
another once they have decohered. Measurements and decoherence represent the
quantum mechanical aspect of time asymmetry (Joos and Zeh, 1985; Gell-Mann and
Hartle, contributions to this conference) that also has to be derived from the
structure of the Wheeler-DeWitt wave function.

A procedure for deriving the approximate concept of a time-dependent wave
function $\psi(t)$ from the Wheeler-DeWitt equation has been proposed by means
of
the WKB approximation (geometric optics) valid for part of the dynamical
variables of the universe. These variables may be those describing the spatial
geometry (Banks, 1985), those forming the `mini superspace' of all monopole
amplitudes on a Friedmann sphere (Halliwell and Hawking, 1985), or all
macroscopic variables which define an appropriate `midi superspace'. For
example, Halliwell and Hawking assumed that the wave function of the universe
can approximately be written as a sum of the form
$$\Psi_{universe}\approx\sum_r e^{iS_r(\alpha,\Phi)}\psi_r(\alpha,\Phi;
\{x_n\}),\eqno(2)$$
where $\alpha = \ln a$ is the logarithm of the expansion parameter, $\Phi$ is
the monopole amplitude of a
massive scalar field which represents matter in this model, while the variables
$x_n$  (with $nJ>J0$) represent all multipole amplitudes of order $n$. The
exponents $S_r(\alpha,\Phi)$ are Hamilton-Jacobi functions with appropriate
boundary conditions, while the relative  states
$\psi_r$ are assumed to depend only weakly on $\alpha$ and $\Phi$. If the
corresponding orbits of geometric optics in mini superspace are parametrized in
the form $\alpha(t_r),\Phi(t_r)$, one may approximately derive from the
Wheeler-DeWitt equation a Schr\"odinger type evolution
 $$i{\partial\over{\partial t_r}}\psi_r(t_r,\{x_n\})=H_x\psi_r
(t_r,\{x_n\})\eqno(3)$$
for the `relative states'
$\psi_r(t_r,\{x_n\}) := \psi_r(\alpha(t_r),\Phi(t_r),\{x_n\})$. It may apply
within the limits of geometric optics along most parts of the trajectories on
each WKB sheet $S_r(\alpha,\Phi)$, but one must keep in mind that this
dynamical approximation does not {\it define} the states $\psi_r(t_r,\{x_n\})$
from which the entropy is to be calculated.

In order to be able to describe the dynamics of the
observed quantum world, equation (3) must
contain the description of the above-mentioned measurements and
measurement-like
interactions in von Neumann's unitary form
$$\psi_r\propto\left(\sum_kc_k\psi^S_k\right)\psi^A_0\Buildrel\to\over{t_r}
\sum_kc_k\psi^S_k\psi^A_k , \eqno(4)$$
valid in the direction of `increasing time'. For
proper measurements the  `pointer positions' $\psi^A_k$ of the `apparatus' $A$
must decohere through further `measurements' by the environment, and thus lead
to newly separated world branches, each one with its own corresponding
`conditioned (physical) entropy'. The formal entropy corresponding to the
ensemble of different values of $k$ would instead have to be interpreted as
describing `lacking knowledge'.

This required asymmetry with respect to the direction of the orbit parameter
$t_r$  means that (3) may be meaningfully integrated, starting from the wave
function representing the present state of the observed world, only into the
`future' direction of $t_r$ (where it has to describe the entangled
superposition of all outcomes of future measurements). In the `backward'
direction of time this calculation does {\it not} reproduce the correct quantum
state, since the unitary predecessors of the non-observed components would be
missing. This is particularly important if the trajectories are continued
backwards into the inflationary era, or even into the Planck era where
different
trajectories in mini superspace (and in the case of recollapsing universes even
both of their `ends') have to interfere with one another in order to form the
complete boundary condition for the total Wheeler-DeWitt wave function (the
`intrinsic' initial condition).

Entropy is expected to grow in the same direction of time as that
describing measurements. Any such asymmetry requires a {\it very special cosmic
initial condition}; the existence of measurement-like processes
 in the quantum world requires essentially a non-entangled initial state (Zeh,
1992). Since the unitary dynamics (3) was derived as an approximation from the
Wheeler-DeWitt equation, its initial condition for $\psi_r(t_r)$, too, must
be derived from $\Psi_{universe}$. There are no free boundary
conditions for trajectories or their relative states.

In order to obtain an appropriate asymmetry of the Wheeler-DeWitt wave
function, it will be assumed in accordance with current models of the
quantum universe that the Wheeler-DeWitt Hamiltonian for the gauge-free
multipoles on the Friedmann sphere is of the form
$$2e^{3\alpha}H=+{{\partial^2}\over{\partial\alpha^2}}-{{\partial^2}\over
{\partial\Phi^2}}-\sum_n{{\partial^2}\over{\partial x^2_n}}+V(\alpha,
\Phi,\{x_n\}),\eqno(5)$$
with a potential $V$ that becomes `simple' (e.g. constant) in the limit J
$\alpha\to-\infty$.  In his talk, Julian Barbour gave an example for how
complicated the effective potential in configuration space becomes instead once
the particle concept has emerged from the general quantum state of the
fundamental fields. The hyperbolic nature of (5) defines an initial value
problem
with respect to $\alpha$ which then also allows one to choose a `simple' (or
symmetric) initial condition (SIC) for $\Psi_{universe}$ in the limit of small
$a$. Its qualitative aspects may be illustrated by a WKB
approximation with respect to $\alpha$ (Conradi and Zeh, 1991; Conradi, 1992)
 $$\Psi_{universe}(\alpha,\Phi,\{x_k\})\to{1\over{(-V)^{1/4}}}\exp
\left[\int^\alpha_{-\infty}
\sqrt{-V(\alpha',\Phi,\{x_k\})}d\alpha'\right]\to\Psi(\alpha)\eqno(6)$$
for $\alpha \to -\infty$. The explicit form of the `initial' wave function
resulting from the no-boundary condition (Hartle and Hawking, 1983) is less
obvious, but need not be different from (6).

If the initial simplicity of the relative states $\psi_r$ of (2) can be derived
from this or some similar simple structure of the total wave function close to
the singularity, this means that `early times' (in the thermodynamical sense)
must correspond to small values of $a$. However, {\it classical} trajectories
in
the mini superspace spanned by $a$ and $\Phi$ return to small values of $a$ for
closed universes with cosmological constant $\Lambda \leq 0$ (even though they
are clearly not symmetric in the generic case -- see Fig.~1).\footnote{*}{The
`no-boundary' condition, defined as a boundary condition for the Wheeler-DeWitt
{\it wave function}, is sometimes also used for deriving special `initial'
conditions for {\it trajectories} at {\it one} of their ends, which
are then {\it classically} continued through all of their history (cf. Laflamme
and Shellard, 1987). The required classical conditions at small values of $a$
are
thereby often in violent conflict with the uncertainty relations. However, such
a selection of {\it trajectories}
is neither compatible with the usual probability interpretation of quantum
mechanics, nor with the structure of the Wheeler-DeWitt wave function derived
from its boundary condition. By no means should these trajectories be used to
calculate `corrections' to the wave function from which they were obtained as
approximate and limited concepts. (Classically, the exceptional condition of a
`bounce' at small values of $a$, sometimes derived in this way, would describe
the middle of a universe's history, not its beginning or end.)} How, then, can
one distinguish between the Big Bang and the Big Crunch? Or is that distinction
really required for the definition of an arrow of time in quantum gravity?

\bigskip \bigskip {\bf Fig.~1:} Asymmetric
classical trajectory in mini superspace. (After Hawking and  Wu, 1985 -- see
also
Laflamme, this conference.) $a$ is plotted upwards, $\Phi$ from left to right.
Dotted curve corresponds to $V = - a^4 + m^2 a^6 \Phi ^2 = 0$. In more than
two-dimensional mini superspace, the trajectories need not intersect
themselves. If the corresponding wave packets (Fig.~2) do not even overlap
thereby, this would in reduced dimensions be described as their
decoherence from one another.

 \medskip
The contributions of Murray Gell-Mann, Jim Hartle and Larry Schulman to
this conference indicate that it is not, provided the considered
universe is very young compared to its total lifetime. A symmetric
(double-ended) low entropy condition for an assumed $\Psi_{universe}(t)$ would
be allowed even if the latter obeyed a unitary time dependence (although
it would then represent a very strong constraint). In quantum gravity, however,
where there is no general time parameter $t$, one has to conclude that a
`simple' condition for $\psi_r(t_r)$ can either be derived from the boundary
condition for $\Psi_{universe}$ at both ends of a turning
quasitrajectory in mini superspace, or at none. (Any asymmetric selection
criteria for trajectories or their relative states -- for example by means of a
time-directed probability interpretation -- would {\it introduce} an absolute
direction of time `by hand'.) In the second case, the asymmetry of the world
would have to be explained as a `great accident' occurring at one end. In the
first case, all `statistical' arrows of time must reverse their direction
together with the expansion of the universe. Integrating (3) in the
asymmetric sense of (4) beyond the cosmic turning point would precisely
correspond to {\it presupposing} the quantum mechanical arrow to keep its
direction.

If the concept of trajectories through mini superspace were applicable at all
for this purpose
(cf. however Sect.~2), the derivation of {\it thermodynamically} asymmetric
universes would require the existence of two extremely different regions at
small values of $a$, together with a proof that {\it almost all} trajectories
compatible with the structure of the correct Wheeler-DeWitt wave function have
one of their ends in each of them. This would not only seem to be in conflict
with the sensitivity of the trajectories to their initial conditions, but also
with the statistical interpretation of entropy. In correct quantum description,
the broad `initial wave packet' which represents the whole assumed low entropy
region would, if exactly propagated through superspace according to the
Wheeler-DeWitt equation and reflected from the repulsive curvature potential at
large $a$, have to reproduce the complementary `initial' wave packet that
represents the high entropy region at the boundary of small $a$ without thereby
interfering with the low entropy region. The condition of reflection
(integrability for $a \rightarrow \infty$) restricts the otherwise complete
freedom of choosing the {\it intrinsic} initial values (corresponding to the
hyperbolic nature of the Wheeler-DeWitt equation) by a factor of $1/2$.
Regardless of all open problems of dynamical consistency, no properties of the
Wheeler-DeWitt Hamiltonian or in the no-boundary condition seem to indicate the
existence of two that much contrasting regions for small values of $a$.

If the arrow of time is instead correlated with the expansion, the derived
dynamics (3) for $\psi_r$ has always to be applied in the
direction of growing values of $a$. In particular, considering the inflation of
the early universe as `causing' a low entropy state at one end of the
trajectory only would be equivalent to {\it presuming} an arrow of causality in
a certain direction of it (instead of deriving this asymmetry as claimed).

Notice that in quantum gravity there is no problem of consistency of
the lifetime of the recollapsing universe with its supposedly much longer
Poincar\'e cycles (that is, with the mean time intervals between two {\it
statistical} fluctuations of cosmic size), as it would arise from the
mentioned double-ended boundary conditions under deterministic (such as
unitary)
dynamics. The exact dynamics $H\Psi_{universe} = 0$, understood as an intrinsic
initial value problem in the variable $\alpha$, constitutes a well-defined
one-ended condition, while the reversal of the arrows of time described by the
time-dependence $\psi_r(t_r)$ is facilitated by the required {\it corrections}
to
the derived unitary dynamics. These corrections have to describe
\underbar{re}coherence and inverse branchings on the return leg.

I am thus trying to convince Stephen Hawking that he did {\it not} make a
mistake\footnote{*}{The title of Hawking's presentation at the conference
was ``My greatest mistake''.} before he changed his mind about the arrow of
time! Even in classical general relativity, the asymmetry of individual
trajectories in mini superspace (pointed out by Don Page, 1985) would not be
sufficient for drawing conclusions on much stronger {\it thermodynamical}
asymmetries.
\bigskip\noindent
{\bf 2 Reversal of the expansion of the
universe in quantum gravity}
\medskip
Within the canonical quantum theory of gravity it appears therefore
hardly {\it possible}  for the arrow of time to maintain its direction when the
universe starts recollapsing. However, the above picture of wave functions
approximately evolving along separate WKB orbits in mini superspace is not a
sufficient representation of the dynamics described by the
Wheeler-DeWitt equation -- not even far outside the Planck region. As will be
shown, the approximation of geometric optics does not justify the continuation
of classical trajectories through the whole history of a universe. For example,
a trajectory chosen to be compatible with the WKB approximation of the wave
function at one end, and found to be incompatible with it at the other one,
would {\it not} indicate an asymmetric arrow of time along this
trajectory, but simply demonstrate that the concept of trajectories must have
broken down in between.

Wave mechanically, trajectories have to be replaced by narrow wave packets
which
separately solve the wave equation. The exact
dynamics for $\Psi_0(\alpha,\Phi)$ in mini superspace
(now replacing the approximation
$e^{iS(\alpha,\Phi)}$) is described by
$$2e^{3\alpha}H\Psi_0(\alpha,\Phi)={{\partial^2\Psi_0}\over{\partial\alpha^2}}
-{{\partial^2\Psi_0}\over{\partial\Phi^2}}+[-e^{4\alpha}+m^2e^{6\alpha}\Phi^2]
\Psi_0(\alpha,\Phi)=0.\eqno(7)$$
The $\alpha$-dependent oscillator potential for
$\Phi$ suggests the  ansatz
$$\Psi_0(\alpha,\Phi)=\sum_n c_n(\alpha)\Theta_n\left(\sqrt{me^{3\alpha}}
\Phi\right),\eqno(8)$$
where the functions $\Theta_n$ are the oscillator eigenfunctions. In
the adiabatic  approximation, the coefficients $c_n(\alpha)$ decouple
dynamically,
$${{d^2c_n(\alpha)}\over{d\alpha^2}}+[-e^{4\alpha}+(2n+1)me^{3\alpha}]
c_n(\alpha)=0.\eqno(9)$$
In this case, coherent oscillator wave packets exhibit
the least possible  dispersion, and may therefore be expected to resemble the
trajectories of geometric optics best.
\bigskip
\bigskip
     {\bf Fig.~2:} Wave packet representing the trajectory of an expanding
universe
(first cosine of Eq. (10) only) for mass of scalar field $m = 0.2$ and mean
excitation $\bar n = 600$, corresponding to $a_{max} = 240$. Plot range from
left
to right is $-0.19 < \Phi < 0.19$, while from bottom to top it is $50 < a <
150$.
(The intrinsic structure of the wave packet is not resolved by the chosen grid
size.)
\medskip
 As demonstrated by Kiefer (1988), the usual (here `final' with respect  to the
intrinsic wave dynamics) condition of square integrability for
$\alpha\to+\infty$
leads to the classically expected reflection of quasitrajectories from the
repulsive curvature-induced potential $- e^{4\alpha}$. (Without such a
condition, wave packets would not return at all.) For example, a further WKB
approximation to (9), together with Langer's pasting to the exponentially
decreasing WKB solutions at the classical turning point, leads to
$$\eqalign{c_n(\alpha)&\propto\cos[\phi_n(\alpha)+n\Delta\phi]\quad
+\quad\cos[\phi_n(\alpha)-n\Delta\phi+\delta_n]\cr &=\hbox{`expanding
universe'}+\hbox{`collapsing universe'},\cr} \eqno(10)$$
where the $\phi_n$'s
are monotonic functions of $\alpha$ (approximately proportional  to $n$) while
$\delta_n=(\pi/4)m^2(2n+1)^2$ is the `scattering' phase shift enforced by the
`final' (large $a$) condition. The two cosines correspond to the expanding and
recollapsing parts of the histories of classical universes in mini
superspace (in a merely relative sense, of course). $\Delta\phi$ is the phase
of
the classical $\Phi$-oscillation at the point of maximum $\alpha$ (describing
the
asymmetry of the trajectory). If the constants of integration at our disposal
from (9), which determine the size and phase of the coefficients $c_n$, are now
chosen to form coherent states from the first cosine on the rhs of Eq. (10),
these phase relations are then completely changed by the large phase shift
differences $\delta_n-\delta_{n-1}\propto n$ resulting from the second cosine.
While the term representing the expanding universe (Fig.~2) nicely resembles
the
corresponding part of a classical trajectory (Fig.~1), the reflected wave is
smeared out over the whole allowed region (Fig.~3). This spreading must also be
described by the corresponding Klein-Gordon current. From a sharp
($n$-independent) potential barrier in `time' $\alpha$, the wave packets would
instead be reflected without any dispersion.

\bigskip \bigskip {\bf Fig.~3:} Same wave packet as in Fig.~2 with recollapsing
part (second cosine) added. The part of the wave packet representing the
expanding universe of Fig.~2 is still recognizable.
\medskip
This dispersion of the wave packet will become even more important for more
macroscopic  universes (higher mean oscillator quantum numbers $\bar n$), since
the phase shift {\it differences} are proportional to $n$. The result depicted
by Fig.~3 may therefore be expected to represent a generic property of
Friedmann
type quantum universes. Quasiclassical trajectories must then never be
continued
beyond the turning point in order to end in a well defined region of high
entropy. The wave mechanical continuation leads instead to a {\it superposition
of many} recollapsing universes (each of which cannot be intrinsically
distinguished from an expanding one). Cosmological quantum effects of gravity
thus seem to be essential not only at the Planck scale! The phase relations of
the resulting superpositions of quasitrajectories on the return leg in mini
superspace are however destroyed by decoherence -- now `irreversibly' acting in
the opposite direction of the trajectory (with increasing $a$ again) because of
the (formally) final condition at the (formal) Big Crunch. (The phase shifts
$\delta_n$ could as well have been put into the first cosine with a negative
sign, since there is no absolute direction of probabilistic `scattering' from
one wave packet into the other. One has to be careful to avoid any notion of
absolute time.) A related result has independently been obtained by Kiefer
(1992b). This is further evidence that the unitary dynamics (3) cannot be
continued along trajectories beyond the turning point at maximum $a$.

Although wave
packets solving the Wheeler-DeWitt equation in mini superspace can thus be {\it
defined} to be intrinsically asymmetric, they are {\it physically} determined
(as
Everett branches) by their decoherence from one another. Wave packets in the
{\it complete} configuration space (which never decohere, since they do not
possess an environment) are not to describe the whole `quantum world', but
merely
the (limited) causal connections which give rise to the latter's `classical
appearance'.
\bigskip\noindent
{\bf 3 Black-and-white holes}
\medskip
A formal reversal of the arrow of time (in particular if facilitated through
quantum effects near the turning point of the universal expansion) must
drastically affect the internal structure of black holes (Zeh, 1992).
For comparison, consider black holes which would form during the expansion of a
time-asymmetric universe, and which are massive enough to survive the turning
point (cf. Penrose's diagram in Fig.~4). If the arrow of time is now
formally reversed along a (quasi)trajectory through mini or midi superspace in
order to form a quasiclassical time-symmetric universe, black holes cannot
continue `losing hair' any further by radiating their higher multipoles {\it
away} (by means of retarded radiation) when the universe starts recollapsing.
They must instead grow hair by means of the now {\it coherently incoming}
(advanced) radiation that has to drive the matter apart again.
 \bigskip \bigskip
     {\bf Fig.~4:} Time-asymmetric classical universe with a homogeneous Big
Bang
only  (Penrose, 1981).
\medskip
The reversal of all arrows of time has of course
to include the replacement of time-directed `causality' by what would formally
represent a `conspiracy'. A mere reversal of the expansion would not by itself
be able to `cause' a reversal of the thermodynamical or radiation
arrows without simultaneous reversal of the time-direction of this causation.
The
(fork-like) causal structure (see Zeh, 1992) must hence be contained in the
dynamical structure of the universal wave function that results from the
intrinsic initial condition by means of the Wheeler-DeWitt equation. Black
holes must therefore formally disappear as `white holes' during the recollapse
phase of the universe.

This surprising fate of black holes thus seems to become important
only in the very distant future (long after horizons and singularities may be
expected to have formed in their interiors). However, our simultaneity with a
black hole is not well defined because of the time translation invariance of
the
Schwarzschild metric. Fig.~5 shows a spherical black hole in Kruskal-type
coordinates (a modified Oppenheimer-Snyder scenario) after translation of the
Schwarzschild time coordinate t such that the turning point of the universal
expansion is now at $t=0$ (hence also at the corresponding Kruskal time
coordinate $v = 0$). The resulting `black-and-white hole' must then also
exhibit a {\it thermodynamically symmetric} appearance, although it need not be
symmetric in non-conserved microscopic or macroscopic properties (`hair'). If
{\it past} horizons and singularities can in fact be excluded by an appropriate
initial condition at the Big Bang (as it is claimed for the Weyl tensor
hypothesis), the same conclusion must hold in quantum gravity also for {\it
future} horizons and singularities. One may therefore conjecture a completely
singularity-free quantum world (i.e., a wave function vanishing at {\it all}
singularities).
 \bigskip
\bigskip
    {\bf Fig.~5:} `Black-and-white hole' originating from a thermodynamically
active (i.e., non-pathological) collapsing spherical matter distribution, with
the Kruskal time coordinate $v = 0$ chosen to coincide with the time of maximum
size of the universe. If the quantum effects studied in Sect.~2 are
essential, this classical picture is not meaningful itself in the region of
`quantum behaviour' around $v = 0$. Only a probabilistic connection can then
exist between its upper and lower parts.
 \medskip
Fig.~6 shows the same situation as Fig.~5 from our perspective of a young
universe (after a back-translation of the Schwarzschild time coordinate such
that $t_{today} = 0$). From this perspective, the time coordinate $t=t_{turn}$
appears to be very `close' to where one would expect the future horizon to
form.
The `strange' thermodynamical and quantum effects now also
appear to occur close to the horizon, thereby preventing it to form.
 \bigskip
 \bigskip
{\bf Fig.~6:} Same black-and-white hole as in Fig.~5 considered from our
perspective of a young universe.
 \medskip
This reversal of the gravitational
collapse cannot be {\it observed} from a safe distance, although it could be
experienced by suicidal methods within relatively short proper times if a black
hole were available in our neighborhood. If the black-and-white hole is massive
enough, this kind of `quantum suicide' must be quite different from the
classically expected one by means of tidal forces. In a classical picture,
travelling through a black-and-white hole may reduce the proper
distance between the Big Bang and the Big Crunch considerably, but
unfortunately
 we could not
survive as information and memory {\it gaining} systems. This consideration
should at least demonstrate that the classical (Kruskal-Szekeres) continuation
of the Schwarzschild metric beyond the horizon is absolutely doubtful for
thermodynamical and quantum mechanical reasons!

Before Stephen Hawking changed his mind about the time arrow in a recollapsing
universe, he had conjectured (Hawking, 1985) that the arrow is reversed
{\it inside} the horizon of a black hole, since ``it would seem just like the
whole universe was collapsing around one'' (cf. also Zeh, 1983). This
consequence would however not describe the situation in a
thermodynamically time-symmetric universe.

Penrose's black holes, hanging like stalactites from the `ceiling' (the Big
Crunch) in Fig.~4, must now also become symmetric, as shown in Fig.~7.
Black-and-white holes in equilibrium with thermal radiation (as studied by
Hawking, 1976) would instead consist of thermal radiation at both ends. They
would possess no `hair' at all, neither to lose nor to grow. The classically
disconnected upper and lower halves of Fig.~7 should rather be interpreted as
two of the many Everett branches of the quantum universe, each of them
representing an {\it expanding} quasiclassical world.
 \bigskip
\bigskip
{\bf Fig.~7:} Time-symmetric, singularity-free universe with black-and-white
holes together with (small) black or white holes.
 \medskip
The absence of singularities from this quantum universe thus appears to be a
combined thermodynamical and quantum effect. However, one may equivalently
interpret the result as demonstrating that in quantum cosmology the
thermodynamical arrow is a {\it consequence} of the absence of inhomogeneous
singularities -- a generalization (or symmetrization) of Penrose's Weyl tensor
condition.
\medskip
{\bf Acknowledgment:} I wish to thank H.D. Conradi and C. Kiefer for their
critical reading of the manuscript.
\bigskip\noindent
{\bf References}
\medskip
\item{1.} Banks, T. (1985) TCP, Quantum Gravity, the Cosmological Constant,
and all that ... {\it Nucl. Physics} {\bf B249}, 332
\item{2.} Conradi, H.D. and Zeh, H.D. (1991) Quantum cosmology as an initial
value problem. {\it Phys. Lett.} {\bf A154}, 321
\item{3.} Conradi, H.D. (1992) Initial state in quantum cosmology. {\it Phys.
Rev.} {\bf D46}, 612
\item{4.} Halliwell, J.J. (1989) Decoherence in quantum
cosmology. {\it Phys. Rev.} {\bf D39}, 2912
\item{5.} Halliwell, J.J. and Hawking, S.W. (1985) Origin of structure
in the Universe. {\it Phys. Rev.} {\bf D31}, 1777
\item{6.} Hartle, J.B. and Hawking, S.W. (1983) Wave Function of the
Universe. {\it Phys. Rev.} {\bf D28}, 2960
\item{7.} Hawking, S.W. (1976) Black
Holes and Thermodynamics. {\it Phys. Rev.} {\bf D13}, 191
\item{8.} Hawking, S.W. (1985) Arrow of Time in Cosmology. {\it Phys. Rev.}
{\bf D32}, 2489
\item{9.} Hawking, S.W. and Wu, Z.C. (1985) Numerical Calculations of
Minisuperspace Cosmological Models. {\it Phys. Lett.} {\bf 151B}, 15
\item{10.} Joos, E. and Zeh, H.D. (1985) The Emergence of Classical Properties
through Interaction with the Environment.  {\it Z. Phys.} {\bf B59}, 223
\item{11.} Kiefer, C. (1987) Continuous measurement of mini-superspace
variables by higher multipoles. {\it Class. Qu. Gravity} {\bf 4}, 1369
\item{12.} Kiefer, C. (1988) Wave packets in mini-superspace. {\it Phys. Rev.}
{\bf D38}, 1761
\item{13.} Kiefer, C. (1992a) Decoherence in Quantum Cosmology. In {\it
Proceedings of the Tenth Seminar on Relativistic Astrophysics and Gravitation,}
Eds. S. Gottl\"ober, J.P. M\"ucket and V. M\"uller, World Scientific
\item{14.} Kiefer, C. (1992b) Decoherence in quantum electrodynamics and
quantum gravity. Phys. Rev. D46, 1658
 \item{15.} Laflamme, R. and Shellard,
E.P.S. (1987): Quantum cosmology and  recollapse. Phys. Rev. D35, 2315
 \item{16.} Page, D.N. (1985) Will entropy increase if the
Universe  recollapses? {\it Phys. Rev.} {\bf D32}, 2496
\item{17.} Page, D.N. and Wootters, W.K. (1983) Evolution without Evolution:
Dynamics Described by Stationary Observables. {\it Phys. Rev.} {\bf D27},
2885
\item{18.} Penrose, R. (1981) Time Asymmetry and Quantum Gravity, In {\it
Quantum Gravity 2}, Eds. Isham, C.J., Penrose, R. and Sciama, D.W.,
Clarendon Press.
\item{19.} Zeh, H.D. (1983) Einstein Nonlocality, Space-Time Structure,
and Thermodynamics, In {\it Old and New Questions in Physics, Cosmology,
Philosophy, and Theoretical Biology}, Ed. van der Merwe, A., Plenum.
\item{20.} Zeh, H.D. (1992) {\it The Physical Basis of the Direction of Time},
Springer (second edition)
\bigskip\noindent
{\bf Discussion}
\medskip
Hawking: Your symmetric initial condition for the wave function is wrong!

Zeh: Do you mean that it does not agree with the no-boundary condition?

Hawking: Yes.

Zeh: It was not meant to agree with it, although we found it to
be very similar to the explicit wave functions you gave in the literature for
certain regions of mini superspace. This is however not essential for my
argument. It requires only that the multipole wave functions $\psi_r$ become
appropriately `simple' (low-entropic and factorizing) for small values of $a$
(as
you too seem to assume, although only at that `end' of the trajectory where you
start your computation).

Barbour: Did I understand you correctly to say that the
criteria Kiefer used to obtain his solution was of the kind I call
Schr\"odinger
type, namely that there should be no blowing up of the wave function anywhere
in
the configuration space?

Zeh: Yes -- if by blowing-up solutions you mean the
exponentially increasing ones. Otherwise you would not be able
to describe reflection (turning trajectories) by means of wave packets. I think
this assumption corresponding to the usual normalizability is natural (or
`naive'
according to Karel Kucha\v r) if the expansion parameter $a$ is considered as a
dynamical quantum variable (as it should in canonical quantum gravity).

Barbour: Could it be that worries about the turning point are an
artifact of the extreme simplicity of the model? Consider in contrast a
two-dimensional oscillator in a wave packet corresponding to high angular
momentum!

Zeh: The described quantum effects at the turning point are due to
the specific Friedmann potential with an oscillator constant for $\Phi$
exponentially increasing with $\alpha$. They do not seem to disappear if added
degrees of freedom possess similarly `normal' potentials (e.g. polynomials
multiplied by positive powers of $a$). This seems to be the case in
Friedmann-type models.

Kucha\v r: Did you study decoherence between $\psi$'s
corresponding to one $S$, or also the decoherence corresponding to different
$S$'s?

Zeh: I expect decoherence to become effective (with increasing $\alpha$)
between
different trajectories in mini superspace (cf. Kiefer, 1987), between
macroscopically different branches of the multipole wave functions $\psi_r$
along every trajectory, and between different WKB sheets corresponding to
different $S$'s except at very small and large $a$ (cf. Halliwell, 1989;
Kiefer,
1992a). Otherwise equation (3) would not be valid as an independent
approximation on different sheets.

Griffiths: In applying ordinary quantum mechanics to a closed system, I do not
know how to make sense out of the `wave function of the closed
system'. I need the unitary transformations that take me from one time to
another. Is there any analogy of this in quantum gravity? For if not, it is
hard
to see how quantum gravity can be used to produce a sensible description of
something like the world we live in.

Zeh: Your question seems to apply to
quantum gravity in general. I think that it is sufficient for the wave function
of the universe to contain correlations between all physical variables --
including those describing clocks. In classical theory these correlations would
be essentially unique, since they would be represented by the trajectories in
the
complete configuration space which remain after eliminating any (physically
meaningless) time parameter. In quantum theory there are no trajectories that
could be parametrized. These quantum correlations must of course obey
`intrinsic'
dynamical laws as they are described by the Wheeler-DeWitt equation. From them
one tries to recover the time-dependent Schr\"odinger equation (which has to
describe the `observed world') as an approximation when spacetime (the history
of spatial geometry) is recovered as a quasiclassical concept.

Lloyd: Could you
clarify how black holes would grow hair in the contraction phase? Is it through
interference between incoming radiation and the Hawking radiation?

Zeh: Only
the advanced radiation is essential, since black holes can form by
losing hair even if Hawking radiation is negligible. This is a pure symmetry
consideration. A final condition which is thermodynamically and quantum
mechanically (although not in its details) the mirror image in time
of an initial condition that leads to black holes must consequently lead to
their time-reversed phenomena. If Hawking radiation is essential (as for small
mass), the black hole may disappear before $t(a_{max})$ is reached, but again
before an horizon forms.

Hawking: The no-boundary condition can only be interpreted by
means of semi-classical concepts such as the saddle point method.

Zeh: I would
prefer to understand such a fundamental conclusion as the arrow of time in
terms
of an exact (even though incomplete) description. In particular,
your opposite conclusion about the arrow of time seems to be introduced by the
direction of computation (along the assumed trajectories) by using
approximations, similar to how it is often erroneously argued in the theory
of chaos by using `growing errors' in the calculation for explaining the
increase of `real' physical entropy!

 -- Did I understand you correctly during your talk that you -- at the time
when you made what you call your `mistake' -- also expected black holes to
re-expand during the recollapse of the universe?

Hawking: Yes. I did not
understand black holes sufficiently until I changed my mind.
\bye